\documentclass[3p,times,preprint,12pt]{elsarticle}

\usepackage{epsfig,float,amssymb,latexsym,amsmath,amsthm,fancyhdr,slashed}
\usepackage{graphics,psfrag,longtable,color}
\usepackage{bm}
\usepackage{color}

\usepackage{mathtools}

\allowdisplaybreaks
\newcommand{\be}{\begin{eqnarray}}
\newcommand{\ee}{\end{eqnarray}}
\newcommand{\beq}{\begin{equation}}
\newcommand{\eeq}{\end{equation}}

\newcommand{\ord}[1]{{\cal{O}}( #1 )}
\newcommand{\B}{{\bf B}}

\renewcommand{\refname}{}
\usepackage{hyperref}
\hypersetup{pdfauthor={me}, colorlinks=true, citecolor=blue, urlcolor=blue, linkcolor=black}
\biboptions{longnamesfirst,angle,semicolon}

\setcitestyle{square}

\begin{document}

\begin{frontmatter}



\title{Baryon masses and $\sigma$ terms in $\rm SU(3)$ $\rm BChPT \times 1/N_c$}
\author{I.~P.~Fernando$^{**}$}
\author{J.~M.~Alarc\'on$^ {*}$}
\author{J.~L.~Goity$^{*,**}$}
\address{* Thomas Jefferson National Accelerator Facility, Newport News, VA 23606, USA.
\\ ** Department of Physics, Hampton University, Hampton, VA 23668, USA.
}
\tnotetext[mytitlenote]{JLAB-THY-18-2679}


\begin{abstract}
  Baryon masses and nucleon $\sigma$ terms are studied with the effective theory that combines the chiral and $1/N_c$ expansions for  three flavors.  In particular the connection between the deviation of the Gell-Mann-Okubo relation and the $\sigma$ term associated with the scalar density $\bar u u+\bar d d-2\bar s s$ is emphasized. The latter is at lowest order related to a mass combination whose low value has given rise to a $\sigma$ term puzzle. It is shown that while the nucleon $\sigma$ terms have a well behaved  low energy expansion, that mass combination is affected by large higher order corrections non-analytic in quark masses. Adding to the analysis lattice QCD baryon masses, it is found that $\sigma_{\pi N}=69(10)$~MeV and $\sigma_s$  has natural magnitude within its relative large uncertainty.
  
 \end{abstract}


\end{frontmatter}


\section{Introduction}

Baryon mass  dependencies on quark masses, quantified by the different $\sigma$-terms, are among the fundamental observables in baryon chiral dynamics. 
In particular, they give information on the baryon matrix elements of scalar quark densities, for which there is no alternative way  for their determination. 
 The definition of $\sigma$ terms is through the Feynman-Hellmann theorem\footnote{The following notation will be used: $\sigma_i(B)=m_i\frac{\partial}{\partial m_i} m_B$, where $m_i$  indicates a quark mass ($i=u,d,s$) or combination thereof ($0,3,8$), and $B$ is a given baryon. When $B$ is not explicitly indicated it is assumed to be a nucleon.},  which, for three flavors, through the physical baryon masses gives access to only two such terms, namely those associated with the SU(3) octet quark mass combinations $m_3=m_u-m_d$ and $m_8=\frac{1}{\sqrt{3}}(\hat m-m_s)$, where $\hat m$ is the average of the $u$ and $d$ quark masses. The $\sigma$ terms associated with the singlet component $m_0=\frac 13(2 \hat m+m_s)$ require knowledge of baryon masses for unphysical quark masses, which is made possible through lattice QCD (LQCD) calculations. On the other hand, the   pion-nucleon $\sigma$ term $\sigma_{\pi N}\equiv \frac{\hat m}{2 m_N} \langle N\mid \bar u u+\bar d d\mid N\rangle$ is accessible through its connection to pion-nucleon scattering via a low energy theorem \cite{GellMann:1968rz,Cheng:1970mx,Brown:1971pn}. Such a determination of  $\sigma_{\pi N}$   had  a long  evolution through the availability of increasingly accurate data and the development of combined methods of dispersion theory and chiral perturbation theory \cite{Gasser:1988jt,Gasser:1990ce,Gasser:1990ap,Pavan:2001wz,Alarcon:2011zs,Hoferichter:2015dsa,Hoferichter:2015hva,Hoferichter:2016ocj}. The values obtained for  $\sigma_{\pi N}$ range from $\sim45$~MeV   \cite{Gasser:1988jt,Gasser:1990ce,Gasser:1990ap} to $\gtrsim 58$~MeV \cite{Pavan:2001wz,Alarcon:2011zs,Hoferichter:2015dsa,Hoferichter:2015hva,Hoferichter:2016ocj,RuizdeElvira:2017stg}, where the difference between the results of the different dispersive analyses resides mostly  in  the different values  of the S-wave $\pi N$ scattering lengths $a^{1/2,3/2}$ used in the subtractions, cf. \cite{RuizdeElvira:2017stg}.
In addition to the results from the analyses of $\pi N$ scattering,  LQCD calculations     extrapolated to or at  the physical point obtain different results, with  values consistent with the recent    $\pi N$  results \cite{WalkerLoud:2008bp} and  smaller,   $\sigma_{\pi N} \approx 40$~MeV \cite{Durr:2015dna, Yang:2015uis, Abdel-Rehim:2016won, Bali:2016lvx}.  The relatively large range of values obtained for $\sigma_{\pi N}$ keeps it as an active topic of study, and in part motivates the present work. An additional motivation is   the relevance of scalar quark operator matrix elements,    quantities that are relevant in studies of direct dark matter detection \cite{Bottino:1999ei, Bottino:2001dj, Ellis:2008hf},  and of  lepton flavor violation  through $\mu-e$ conversion in scattering with nuclei \cite{Cirigliano:2009bz}.

 A puzzle that has been emphasized for a long time \cite{Leutwyler:2015jga} is the relation between $\sigma_{\pi N}$ in the isospin symmetry limit and  the nucleon's $\hat \sigma\equiv  \sqrt{3} \frac{\hat m}{m_8} \sigma_8$, namely $\sigma_{\pi N}=\hat\sigma+ 2\frac{\hat m}{m_s} \sigma_s$, which    for  a natural size value of $\sigma_s$ should give $\sigma_{\pi N}\sim\hat\sigma$. The origin of the puzzle is the relation: $\sigma_8=\frac 13(2 m_N-m_\Sigma-m_\Xi)$ (or other combinations related via the Gell-Mann-Okubo (GMO) relation) valid at linear order in quark masses, which gives $\hat \sigma\sim 25$~MeV. If that relation is a reasonable approximation to the value of $\hat \sigma$, the implication is that, contrary to expectations,  $m_s$ must give a very large contribution to the nucleon mass even for the smaller values of $\sigma_{\pi N}$.  The puzzle is particularly striking for the larger values that have been obtained for   $\sigma_{\pi N}$, which would imply $\sigma_s\sim 0.5$ GeV!. Indeed, this is clearly impossible if one considers that    $\sigma_s=\ord {\frac{1}{N_c}} \sigma_{\pi N}$. 
 
 This work analyzes the $\sigma$ terms    through  the octet and decuplet  baryon masses in the combined chiral and $1/N_c$ expansions $\rm BChPT\times 1/N_c$. 
 The   emphasis   is in that  the effective theory can give at NNLO (one chiral loop) a natural description of   baryon masses, including LQCD results, along with the axial couplings which have been obtained in LQCD at different quark masses. In particular, the resolution of the $\sigma$ term puzzle is explained by the fact that   $\Delta \sigma_8\equiv\sigma_8-\frac 13(2 m_N-m_\Sigma-m_\Xi)$   receives  large non-analytic in quark mass corrections dominated by $m_s$. It will also be shown that $\sigma_8$ itself, and thus $\hat \sigma$, has a natural low energy expansion and therefore the origin  of the puzzle resides in   the large non-analytic correction to the mass combination $\frac 13(2 m_N-m_\Sigma-m_\Xi)$. In fact, a big part of that large correction stems from the contribution of decuplet baryons in the loop, as it was found in Refs. \cite{WalkerLoud:2008bp,Alarcon:2012nr}.
 By analyzing   LQCD baryon masses \cite{Alexandrou:2014sha}, it is found that as expected $\sigma_{\pi N}\sim\hat\sigma$, with the results $\sigma_{\pi N}=69(8)(6)$ MeV,  where the errors are respectively the statistical and theoretical (expected NNNLO corrections) ones, and $\mid\sigma_s\mid\lesssim 50$ MeV.  
  The connection between the deviation from the GMO relation, $\Delta_{GMO}\equiv 3m_\Lambda+m_\Sigma-2(m_N+m_\Xi)$, and $\Delta \sigma_8$,   both calculable at NNLO and given solely in terms of non-analytic loop contributions, is of particular importance in the present work.

\section{ $\rm BChPT\times 1/N_c$ analysis of masses and $\sigma$ terms}

The combined  $\rm BChPT\times 1/N_c$ \cite{Jenkins:1995gc,FloresMendieta:1998ii,FloresMendieta:2012dn,CalleCordon:2012xz,Fernando:2017yqd} implements the consistency of the effective theory with both the approximate chiral symmetry and the expansion in $1/N_c$ of QCD. The expansion requires a link between the chiral and the $1/N_c$ expansions: in practice the natural  link is the $\xi$ expansion where $\ord{p}=\ord{1/N_c}=\ord{\xi}$, which is closely related to the so called small scale expansion \cite{Hemmert:1996xg,Hemmert:1997ye} even when that one did not strictly implement the constraints of the $1/N_c$ expansion. Consistency with $1/N_c$ power counting demands the imposition of a dynamical  SU(6) spin-flavor symmetry, which is broken by sub-leading corrections in $1/N_c$ and   requires the inclusion of the higher spin baryons (the decuplet in the case $N_c=3$) and relates  low energy constants (LECs) in the chiral Lagrangian.  The details on the calculations of baryon masses concerning the present work can be found in \cite{Fernando:2017yqd}. 

The chiral Lagrangian to $\ord{\xi^3}$, including electromagnetic corrections to the baryon masses is given by \cite{Fernando:2017yqd}:

\begin{align}\label{Eq:Lagrangians}
{\cal{L}}_B&=\B^{\dagger}\left(i D_0 +  \mathring{g}_A
	u^{ia}G^{ia}-\frac{C_{HF}}{N_c}{\hat { {S}}^2} - \frac{1}{2 \Lambda} c_2 \hat\chi_+ + \frac{c_3}{N_c\,\Lambda^3} \;\hat\chi_+^2  \right. \nonumber \\
	&  \left.    + \frac{h_1}{N_c^3} \hat S^4+ \frac{h_2}{N_c^2\Lambda} \hat\chi_+\hat{S}^2 + \frac{h_3}{N_c\Lambda} \chi^0_+\hat{S}^2 + \frac{h_4}{N_c\, \Lambda} \; \chi_+^a\{S^i,G^{ia}\} +\alpha \hat Q+\beta \hat Q^2\right) \B.
\end{align}
where terms not directly relevant to the baryon masses have been omitted. 
$M_0=\ord{N_c}$ is the spin-flavor singlet piece of the baryon mass that provides the large mass expansion parameter for HBChPT. In addition to the well known chiral building blocks,  $\B$ represents the baryon spin-flavor multiplet field, $\hat S^2$ is the square of the baryon spin operator, $G^{ia}$ are the spin-flavor generators of SU(6), and $\hat Q$ is the electric charge operator. No   baryon-spin dependent electromagnetic effects are included.   The term proportional to $C_{HF}$ gives the leading order mass splitting between the spin 1/2 and 3/2 baryons. $\mathring g_A$ is identified with $\frac 65 g_A^N$ at the LO, whose physical value is $1.2723\pm0.0023$.  The term $h_1$ is only relevant if baryons with higher spin than 3/2 appear, which requires $N_c\geq 5$. The rest of the terms describe the quark mass effects.
The combination $\hat\chi_+ = N_c \;\chi_+^0 + \tilde\chi_+ $, where $\rm \chi_+^0=\frac 13 Tr \;\chi_+$ and  $\tilde \chi_+$ is the traceless piece of $\chi_+$,    assures that the nucleon mass dependency on $m_s$ is at most $\ord{N_c^0}$ (OZI). $\Lambda $ is an arbitrary scale, which is conveniently chosen to be $m_\rho$.
The baryon mass formula then reads (neglecting isospin breaking for now)\cite{Fernando:2017yqd}:
\begin{align}\label{Eq:MassFormula}
m_B&= M_0+\frac{C_{HF}}{N_c} \hat S^2-\frac {c_1}{ \Lambda} 2B_0(  \sqrt{3}m_8 Y+N_c m_0) 
-\frac{c_2}{\Lambda} 4B_0 m_0-\frac{c_3}{N_c \Lambda^3} \left( 4B_0( \sqrt{3} m_8 Y+N_c m_0)\right)^2\nonumber\\
&- \frac{h_1}{N_c^2 \Lambda} \hat S^4-\frac{h_2}{N_c \Lambda}4B_0( \sqrt{3} m_8 Y+N_c m_0)\hat S^2- \frac{h_3}{N_c \Lambda} 4B_0 m_0 \hat S^2\nonumber\\
&-\frac{h_4}{N_c \Lambda}  \frac {4B_0 m_8}{\sqrt{3}}\left( 3 \hat I^2-\hat S^2-\frac{1}{12} N_c(N_c+6)+\frac{1}{2} (N_c+3)Y-\frac{3}{4} Y^2 \right)+\delta m_B^{\rm loop},
\end{align}
where $\delta m_B^{\rm loop}$ can be obtained with some work using the results in \cite{Fernando:2017yqd}, where the details on the mass renormalization and   results for general $N_c$ can be found.

Setting $c_3=0$ \footnote{The 27-plet SU(3) breaking produced by this term is $\ord{\xi^5}$, and thus for the current purposes it can be neglected}, the terms  analytic in quark masses    in Eqn. (\ref{Eq:MassFormula}) lead to the exact GMO   and Equal Spacing mass relations, which are   unchanged at generic $N_c$.  On the other hand at generic $N_c$     the mass relation for $\sigma_8$ at tree level reads:
\begin{align}\label{Eq:sigma8rel}
 \Delta\sigma_8 &=\sigma_8-\frac 19\left( \frac{5N_c-3}{2} m_N-(2N_c-3)  m_\Sigma-\frac{N_c+3}{2} m_\Xi \right)  ,
\end{align}
  The dominant contributions to   $\Delta_{GMO}$  and $ \Delta\sigma_8$
  are   calculable non-analytic contributions.      $\Delta_{GMO}$  is $ \ord{\xi^4}$ and in large $N_c$ limit it is $\ord{1/N_c}$. On the other hand, $\sigma_8$ is $\ord{\xi}$  and it has a prefactor $N_c$,  and $ \Delta\sigma_8$ is  $\ord{\xi^2}$ also with a prefactor  ${N_c}$. $c_3$ gives a contribution to the  $\Delta_{\rm GMO}$ which is $\ord{\xi^5}$, and to  $\Delta\sigma_8$ at $\ord{\xi^4}$,   both  being beyond the accuracy of the present work. 
$\Delta_{\rm GMO}$\footnote{$\Delta_{\rm GMO}$ corresponds to having removed the EM corrections, otherwise it is denoted by $\Delta^{\rm phys}_{\rm GMO}$}  and  $\Delta \sigma_8$ are thus determined by the meson masses and by the LECs $\mathring g_A/F_\pi$, and $C_{HF}$. $\Delta_{\rm GMO}$ depends rather smoothly on $C_{HF}$, and drives to a large extent the determination of $\mathring g_A/F_\pi$.   One finds the interesting fact that the ratio  $\Delta\sigma_8/\Delta_{\rm GMO}$ ($\sim -13.5$ for $N_c=3$), which is independent of  $\mathring g_A/F_\pi$, is also almost entirely independent of the value of $C_{HF}$ in a very wide range around its actual value. 

The analysis of the physical octet and decuplet baryon masses suffice to  make the main point of this work. In this case, the LECs $c_2$, $c_3$ and $h_1$ are set to vanish, because at the order of the calculation they are redundant (actually $h_1$ is altogether irrelevant unless $N_c\geq 5$). A fit is carried out including strong and electromagnetic isospin breaking. This requires using the meson masses with isospin breaking, which include $\eta-\pi^0$ mixing (required to have a consistent renormalization of the baryon masses) and   the electromagnetic mass shifts where Dashen's theorem is used, which should be sufficient for the current application.  The electromagnetic addition to $\Delta_{\rm GMO}$ is equal to $-\frac 43 \beta$, while the strong isospin breaking has negligible effect, and the electromagnetic contribution to the $p$-$n$ mass difference is equal to  $\alpha+\beta$. The result of the fit to physical masses is shown in Table (\ref{Table:fits}), Fit 1.

\begin{table}[h]
\begin{center}\small
\begin{tabular}{c@{\hspace{2pt}}c@{\hspace{2pt}}c@{\hspace{2pt}}c@{\hspace{2pt}}c@{\hspace{2pt}}c@{\hspace{2pt}}c@{\hspace{2pt}}c@{\hspace{2pt}}c@{\hspace{2pt}}c@{\hspace{2pt}}c}\\\hline\hline 
 &{\large$\frac{\mathring g_A}{F_\pi}$}&{\large$\frac{M_0}{N_c}$}&$C_{HF}$ & $c_1$ & $c_2$& $h_2$&$h_3$&$h_4$&$\alpha$&$\beta$\\[.21cm]
Fit&${\rm MeV}^{-1}$ & MeV  & MeV & &&&&& MeV & MeV \\[5pt]\hline 
1&$0.0126(2) $ &$364(1)$ & $166(23)$ &$-1.48(4) $&$ 0 $ &$ 0$&$0.67(9) $&$0.56(2)$&$-1.63(24) $&$2.16(22) $\\
2 &$0.0126(3)$&$213(1)$&$179(20)$&$-1.49(4)$&$-1.02(5)$&$-0.018(20)$&$0.69(7)$&$0.56(2)$&$-1.62(24)$&$2.14(22)$\\
3&$0.0126^*$&$262(30)$&$147(52)$&$-1.55(3)$&$-0.67(8)$&$0$&$0.64(3)$&$0.63(3)$&$-1.63^*$&$2.14^*$\\
\hline\hline
&   $\Delta^{\rm phys}_{GMO}$ &$\sigma_8$ & $\Delta\sigma_8$ &$\hat\sigma$ & $\sigma_{\pi N}$&$\sigma_s$&$\sigma_3$ &$\sigma_{ u+d}(p-n)$ &&\\[.2cm]
&${\rm MeV} $ & MeV  & MeV&MeV&MeV&MeV&MeV&MeV&&\\[5pt]\hline  
1&$25.6(1.1) $ & $-583(24)$&$-382(13)$&$70(3)(6)$&$-$&$-$&$-1.0(3)$ &$ -1.6(6) $ &&\\
2 &$25.5(1.5)$&$-582(55)$&$-381(20)$&$70(7)(6)$&$69(8)(6)$&$-3(32)$&$-1.0(4)$&$-1.6(8)$\\
3&$25.8^*$ &$-615(80)$&$-384(2)$&$74(1)(6)$&$65(15)(6)$&$-121(15)$&$-$&$-$&&\\
\hline\hline
\end{tabular}
\end{center}
\caption{ Results from fits to baryon masses.  Fit 1 uses only the physical octet and decuplet masses, Fit 2 uses the physical and the LQCD masses  from Ref. \cite{Alexandrou:2014sha}   with  $M_\pi\lesssim 300$ MeV, and Fit 3 uses only those LQCD masses and imposes the value of $\Delta^{\rm phys}_{GMO}$ determined by the physical masses.  The renormalization scale $\mu$ and the scale $\Lambda$ are taken to be equal to $m_\rho$.   $^*$ indicates an input. An estimated theoretical error of 6 MeV is indicated for $\hat \sigma$ and $\sigma_{\pi N}$.}
\label{Table:fits}
\end{table}%
The information given by LQCD, where the baryon masses have been obtained with $M_K$ approximately constant and varying $m_u=m_d$ in a range where $213 {\rm~MeV}<M_\pi<430 {\rm~MeV}$ \cite{Alexandrou:2014sha}, is very useful for testing the effective theory, and necessary for calculating $\sigma_{\pi N}$. Two different fits that include LQCD baryon masses were performed, shown in Table (\ref{Table:fits}). One fit combines the physical and LQCD masses, up to $M_\pi\sim 300$~MeV, and the other uses only LQCD and the physical value of $\Delta_{GMO}$, which is important for controlling the value of $\mathring g_A/F_\pi$. In these fits the LEC $c_2$ which gives the baryon mass dependencies on the singlet quark mass component $m_0$ becomes significant, and its presence is responsible for the significant change in $M_0$ compared to the physical fit. $M_0$ is very precisely determined by the physical masses; Fit 3 shows that it is much less precise if only LQCD masses are used.
The constant $\beta$ can   be estimated by the relation $2 \beta=m_p-m_n-(m_{\Xi^0}-m_{\Xi^-})$, valid to LO in quark masses, which gives $\beta=2.78\pm0.1$~MeV. The   fit indicates that  higher order terms in quark masses   affect   the extraction of $\beta$.  The theoretical error  for $\hat\sigma$ and  $\sigma_{\pi N}$ accounting for  higher order corrections  was estimated by explicitly expanding in $\xi$ and identifying the size of the contributions; the magnitude of the theoretical error  was then estimated to be $\sim 1/3$ the size  of the last term in the expansion.

The observations derived from  the effective theory and from the fits are the following: \\
i) The value of $\mathring g_A/F_\pi$ is to a large extent fixed by $\Delta_{GMO}$, and it corresponds to a value of $g_A^N$ at LO which roughly a factor  0.75 of the physical one; this agrees with what is observed in the analysis of the axial vector couplings  \cite{Fernando:2017yqd}  provided by LQCD calculations  at different values of quark masses \cite{Alexandrou:2016xok}.\\
 ii) The octet baryons contribute 43\% of    $\Delta_{GMO}$, and 33\% of $\Delta\sigma_8$, which shows the importance of the decuplet contributions. \\
  iv) The first fit   determines   $\sigma_8$. Using the natural renormalization scale $\mu=m_\rho$, the different contributions to $\sigma_8$ are primarily given by the terms $c_1$ ($\sim -870$~MeV), $h_4$ ($\sim 110$~MeV) and the loop contributions ($\sim 190$~MeV), where the   latter two  are the NLO contributions. This seems to be a well behaved expansion. On the other hand the mass combination on the RHS of Eqn. (\ref{Eq:sigma8rel}) has the corresponding pattern $-870$ MeV,
  110 MeV and 570 MeV, the latter loop contribution given by the addition of $\Delta \sigma_8\sim 380$~MeV. The NLO terms in the mass combination are very large and tend to cancel the LO one.\\
  v) The correction $\Delta \sigma_8$  becomes quite large for $M_K> 350$~MeV, being about 70\% of $\sigma_8$  for the physical $M_K$. 
  As mentioned earlier, $\Delta \sigma_8$ and $\Delta_{GMO}$ are determined only in terms of $\mathring g_A/F_\pi$, $C_{HF}$ and the meson masses. The ratio $\Delta\sigma_8/\Delta_{GMO}$ does not depend on $\mathring g_A/F_\pi$, and has virtually no dependence on $C_{HF}$. The ratio is also modestly dependent on $M_K$, going from $\sim -11$ to $\sim -14$ when $M_K$ is increased from 200 to 600~MeV.     \\
 vi) The combined fit of physical and LQCD masses, Fit 2,  is compatible with Fit 1; this in   implies that the chiral extrapolation of the LQCD results to the physical case is consistent.   \\
 vii) The fit to only  LQCD masses and imposing the physical  $\Delta_{GMO}$, Fit 3,    serves for a consistency check, which turns out to be quite reasonable. The LQCD masses   do not   describe correctly the hyperfine mass shifts between the octet and decuplet, which is shown in Fig. (1) where the $\Delta$ mass is systematically large, and this is the reason the resulting $C_{HF}$ has some   difference with the other fits. The extrapolation to the physical case turns out to be from 20 to 50 MeV larger than the physical octet masses, but less accurate for the decuplet ones where the $\Delta$ mass, which is the worst case,  comes out to be about 100 MeV larger than the physical one.\\
viii) It is observed that $ \hat\sigma$ and    $\sigma_{\pi N}$ have both a small and approximately linear  dependency on  $M_K$  in a very wide range. This in particular indicates that $\hat m\,  \sigma_s/m_s$ must remain relatively small throughout. \\
ix)  $\sigma_s$ is poorly determined in the present study because the LQCD results are at approximately fixed $m_s$. Its range of values  is  however  in line with the natural expectations.  A LQCD calculation performed with smaller  $M_K$ than the physical one is necessary   to obtain $\sigma_s$ with better precision and also for    understanding  the effective theory in general. 
\\
 x) The results obtained for $\sigma_{\pi N}$ are consistent with the larger values obtained from $\pi N$ analyses \cite{Pavan:2001wz,Alarcon:2011zs,Hoferichter:2015dsa,Hoferichter:2015hva,Hoferichter:2016ocj}. Note however that a more reliable value would require some more accurate and extensive LQCD results.   Fig. (1)  depicts the result for $\sigma_{\pi N}$ from    Fit 2 and its comparison with other results.\\
  xi) The analysis   also  gives  an estimate of the isospin-breaking  $\sigma$ terms $\sigma_3$ and $\sigma_{u+d}(p-n)$. In addition one can extract the separate contributions $\sigma_q(N)$, $q=u,d$, $N=p,n$. The results are the following: $\sigma_u(p)=26.23$~MeV, $\sigma_d(p)=42.42$~MeV, $\sigma_u(n)=23.82$~MeV, $\sigma_d(n)=46.48$~MeV, which checks with $\sigma_{\pi N}=\hat m(\sigma_u/m_u+\sigma_d/m_d)$. The relation $\sigma_u(p)=\sigma_d(n)$  in the isospin symmetry limit is of course satisfied, but the naive quark model relation in the isospin limit $\sigma_u(p)=2 \sigma_d(p)$ is   significantly violated due to   contributions by the SU(2) singlet component of the quark masses.    \\
  xii) Obviously, the discussion can be extended to the rest of the $\sigma$ terms for the different baryons and   their various relations   \cite{Fernando:2017yqd}.\\
  xiii)  One can compare with an analysis in ordinary HBChPT without the decuplet. In that case $\Delta_{GMO}$ requires   $\mathring g_A/F_\pi$ to be significantly larger (corresponding to $g_A^N=1.48$ at LO),  which despite the lack of the decuplet contributions leads  to values of the $\sigma$ terms which are not very different but somewhat larger than the ones obtained here ($\hat\sigma\sim 83$~MeV, $\sigma_{\pi N}\sim 76$~MeV). So, where is the difference?. The answer is simple: in ordinary HBChPT the corrections to the axial currents couplings have large $N_c$ power violating contributions, which   compounded with the larger value of  $\mathring g_A/F_\pi$ required by  $\Delta_{GMO}$ lead to a  failure in describing  the axial couplings obtained in LQCD at different quark masses  \cite{Alexandrou:2016xok}, in particular their observed small quark mass dependencies.\\
  xiv)  Although the approach followed in recent work \cite{Lutz:2018cqo}  should be expected to give a result for $\sigma_{\pi N}$ similar to the one obtained here, it is actually much smaller. It is not clear to the authors whether this may be entirely due to the different set of LQCD data. However, since $\hat\sigma$ is accurately obtained with only the physical masses, the result of  \cite{Lutz:2018cqo} would require a large negative   $\sigma_s$, which seems to be unlikely within the present framework.\\

\begin{figure}
\begin{center}
\epsfig{file=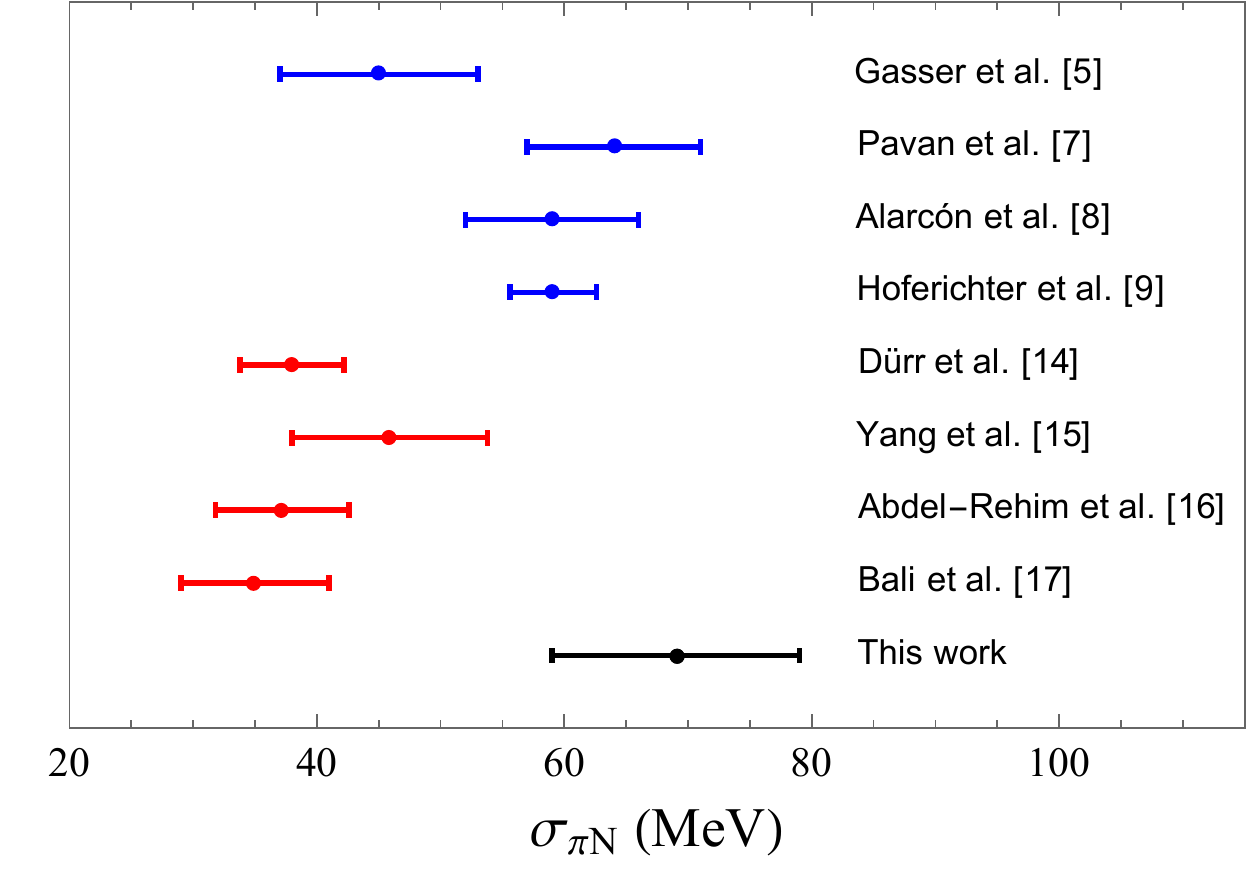,width=0.45\textwidth,angle=0} ~\epsfig{file=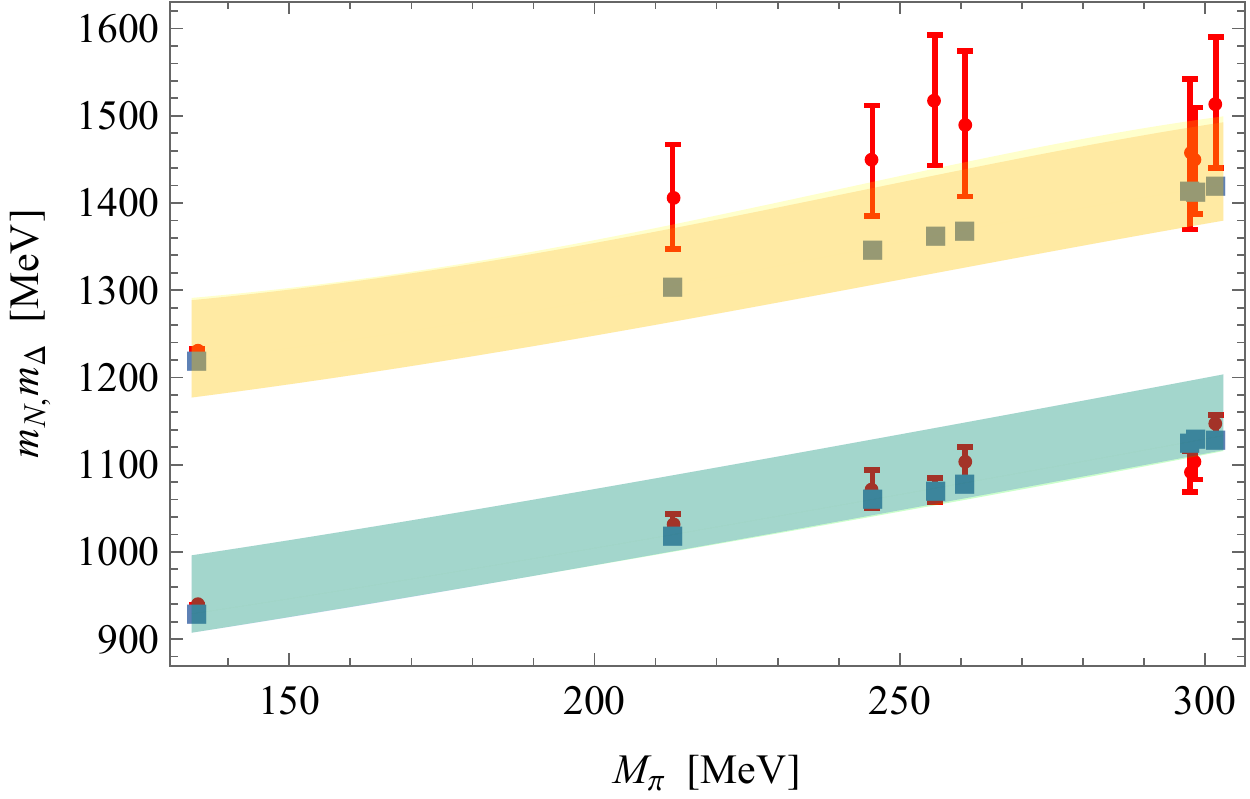,width=0.49\textwidth,angle=0} 
\caption{Left panel: summary of the   determinations of $\sigma_{\pi N}$  from $\pi N$ scattering (blue), from LQCD  (red), and from this work showing the combined  fit and   theoretical error. Right panel: $N$  and $\Delta$ masses from Fit 2 of Table (\ref{Table:fits}): physical and LQCD masses from  \cite{Alexandrou:2016xok}. The squares are the results from the fit and the error bands correspond to 68\% confidence interval.}
\label{Fig:Sigma_terms_figure}
\end{center}
\end{figure}

\section{Summary}

The $\sigma$ terms of nucleons were calculated using  SU(3) ${\rm BChPT \times 1/N_c}$.  From the physical octet and decuplet baryon masses a value of $\hat\sigma$ is obtained which is much larger than the one predicted by a tree level baryon mass combination, in agreement with similar observations in calculations that included the decuplet baryons as explicit degrees of freedom. The "$\sigma$ term puzzle" is understood as the result of large non-analytic contributions to that mass combination, while the higher order corrections to the $\sigma$ terms have natural magnitude. The  intermediate  spin 3/2 baryons play an important role in enhancing $\hat\sigma$ and thus $\sigma_{\pi N}$. The analysis carried out here shows that there is compatibility in the description of $\Delta_{GMO}$ and the nucleon $\sigma$ terms. The value of $\sigma_{\pi N}=69\pm 10$ MeV  obtained here from including LQCD baryon masses agrees with the more recent results from $\pi N$ analyses, where the increase in  value with respect  to previous analyses  has been understood as a result of the values of the input scattering lengths, and strongly disfavor the values from recent  LQCD evaluations.  The tension between results, which   includes LQCD,  remains as an important problem  to which   the present approach  can  hopefully contribute  useful insights. The resolution of that tension will in turn provide a validation test of the approach.

\section*{Acknowledgments}

This work was supported by DOE Contract No. DE-AC05-06OR23177 under which JSA operates the Thomas Jefferson National Accelerator Facility.
JMA  acknowledges partial  support from the MINECO (Spain) and the ERDF (European Commission) through grant No. FPA2016-77313-P. IPF and JLG acknowledge partial support through National Science Foundation through grants  PHY-1307413 and PHY-1613951.




\renewcommand\refname{}
\biboptions{sort&compress }
\setcitestyle{square}
\bibliography{Refs}

%
  
\newpage
%
\end{document}